%% file: joshi.tex
\documentclass[]{raa}
\usepackage{amssymb,amsmath}            % cleaned version: for submission
\usepackage{graphicx,times}
\usepackage{natbib}
\usepackage{color}

\providecommand{\bm}[1]{\mbox{\bol
dmath$#1$\unboldmath}}

\begin{document}

   \title{Long-term photometric study of a faint W UMa binary in the direction of M31}
   
 \volnopage{ {\bf 2017} Vol.\ {\bf X} No. {\bf XX}, 000--000}
   \setcounter{page}{1}

   \author{Y. C. Joshi\inst{1}, Rukmini J.\inst{2}
\thanks{email: \texttt{yogesh@aries.res.in}}}

   \institute{ Aryabhatta Research Institute of Observational Sciences, Manora Peak, Nainital, India - 263002\\
        \and
        Center for Advanced Study in Astronomy, Osmania University, India \\
\vs \no
   {\small Received 2017 ; accepted }
}

\abstract{We carry out a re-analysis of the photometric data in $R_cI_c$ bands which was taken under the Nainital Microlensing Survey during 1998 to 2002 with the aim to detect gravitational microlensing events in the direction of M31. Here, we do photometric analysis of a faint W UMa binary $CSS\_J004259.3+410629$ identified in the target field. The orbital period of this star is found to be 0.266402$\pm$0.000018 day. The photometric mass ratio, $q$, is found to be $0.28\pm0.01$. The photometric light curves are investigated using the Wilson-Devinney (WD) code and absolute parameters are determined using empirical relations which provide masses and radii of the binary as $M_1 = 1.19\pm0.09 M_\odot$, $M_2 = 0.33\pm0.02 M_\odot$  and $R_1 = 1.02\pm0.04 R_\odot$, $R_2 = 0.58\pm0.08 R_\odot$ based on the $R_c$ band data. Almost similar values are found by analysing $I_c$ band data. From the photometric light curve examination, the star is understood to be a low mass-ratio over-contact binary of A-subtype with a high fill-out factor of about 47\%. The binary system is found to be located approximately at a distance of $2.64\pm0.03$ kpc having a separation of $2.01\pm0.05$ $R_\odot$ between the two components.
}

   \authorrunning{Joshi et~al. }            %author_head in even pages
   \titlerunning{Photometric study of W UMa binary towards M31}  % title_head in odd pages
   \maketitle
   
\section{Introduction}
The Nainital Microlensing Survey (NMS) project was conducted for four years during 1998-2002, towards the disk of M31. The project had been carried out with the aim of detecting gravitational microlensing events in the direction of Andromeda galaxy (M31) although data was also found to be well suited for the detection of variable stars in the target field (Joshi et al. 2001). Several surveys aimed to find microlensing events towards M31 have already unearthed thousands of variable stars as a major by-product, most of which were previously unknown (Ansari et al. 2004, Darnley et al. 2004, Fliri et al. 2006, Lee et al. 2012, Soszy\'{n}ksi et al. 2016). Apart from detecting microlensing events in the NMS data (Joshi et al. 2005), we have detected Cepheids (Joshi et al. 2003, 2010) and classical novae (Joshi et al. 2004) in M31. In continuation of our efforts to identify variables in the archival data taken under the survey, we searched for eclipsing binary stars in the target field of M31. As most stars are believed to be in multiple systems, any information gathered on binary stars is vital for understanding stellar evolution and for testing evolutionary models. In this paper, we study a W UMa eclipsing binary in some detail from the multi-band photometric data acquired under the NMS survey.

Most of the W UMa binaries consist of solar-like components having orbital periods in the range of 0.2 to 0.7 days (Qian et al. 2017) and both components of the binary system share a common convective envelope that is placed between the inner and outer critical surfaces in the Roche model. They can be identified by continuous brightness variations and nearly equal depth for primary and secondary eclipses. These stars are ideal candidates to study the formation and evolution scenario in close binaries and provide valuable information on the late stage of stellar evolution that gives the information on the process of mass transfer, angular momentum loss and merging of stars (Li et al. 2014; Yang \& Qian 2015).

The paper is organized as follows: in Section~\ref{dat}, we give the information about the data used in the present analysis. The photometric parameters are determined in Section~\ref{phot}. The light curve analysis is carried out in Section~\ref{analysis} followed by discussion in Section~\ref{discuss}. Our results are summarized in Section~\ref{conclu}\\
\input{table01.tex}
%
%------------------------------------------- Fig01-------------------------
\begin{figure*}[h]
\centering
\includegraphics[angle=0,width=13cm, height=12.0cm]{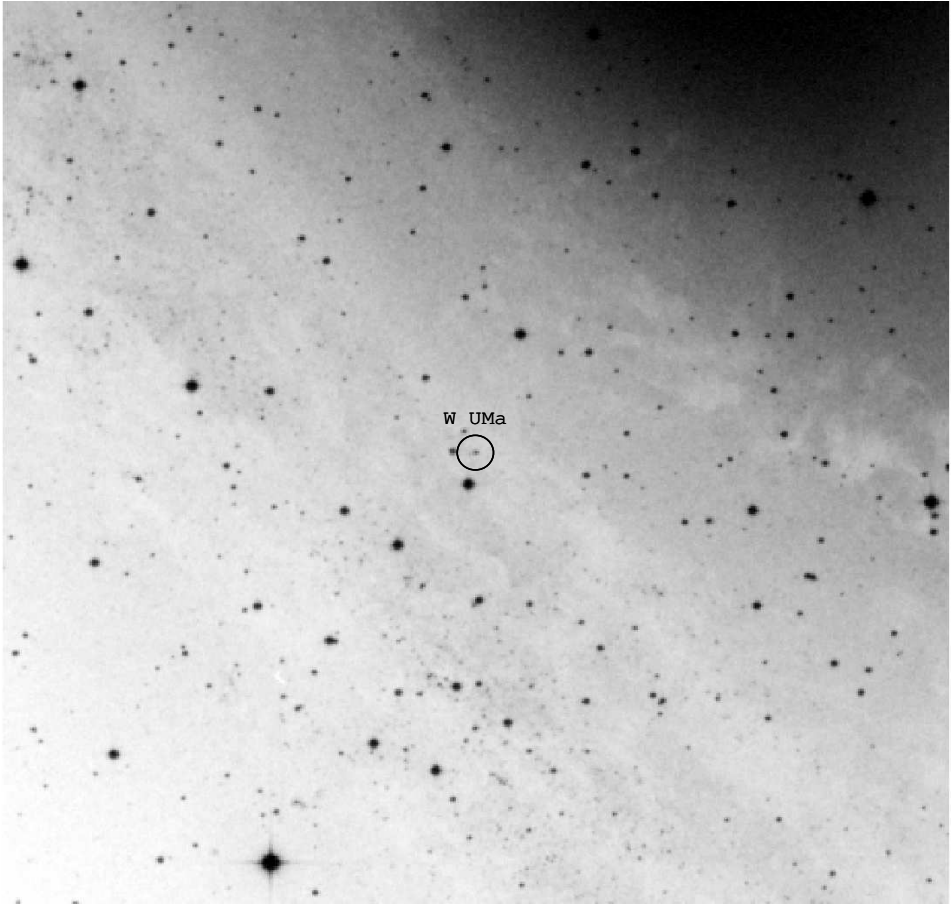}
\begin{flushleft}
\vspace{0.5cm}
\caption{A $13' \times 13'$ finding chart centered at (0:42:59.33, +41:06:29.3). North is up and East is to the right. The W UMa binary is marked in a circle.}
\label{chart}
\end{flushleft}
\end{figure*}
%------------------------------------------- Fig01-------------------------

%
\section{The Data}\label{dat}
The Cousins $R$ and $I$ band (hereafter $R_c$ and $I_c$, respectively) photometric observations of the target field centered at $\alpha _{2000}$ = $0^{h} 43^{m} 38^{s}$, $\delta_{2000}$ = $+41^{\circ}09^{\prime}06^{\prime\prime}$, were carried out for four years in the direction of M31. Starting in November 1998, the survey program continued until January 2002. The observations were started with a field of view of only $\sim 6\times6$ arcmin$^2$ in 1998 but later it was increased to $\sim 13\times13$ arcmin$^2$ in the following years. Although intense observations under the NMS  project were stopped after the observing season 2001-2002 but we continued to acquire the data in the following years whenever telescopes were available. Our overall data set until now consists of 741 images in the $R_c$ and 589 images in the $I_c$ band. A sample log of our observations is given in Table~1 and full table is available through electronically. All the images were processed using IRAF\footnote{Image Reduction and Analysis Facility (IRAF) is distributed by the National Optical Astronomy Observatories, which are operated by the Association of Universities for Research in Astronomy, Inc., under cooperative agreement with the National Science Foundation.}. Standard stars of Landolt's field (1992) were observed and analysed to convert instrumental magnitudes to standard magnitudes. It should be noted that not all the stars were detected in all the frames as observations were taken over a span of many years as well as in different observing conditions.
\section{Photometric Analysis}\label{phot}
Unlike our earlier study (Joshi et al. 2003, 2005, 2010), we have not combined the frames on a nightly basis in the present analysis but carried out photometry on each individual frame in order to increase time resolution. After calibrating the frames, the data were searched for short-period eclipsing binaries, as their evolution is poorly understood. From a visual inspection of bright stars having good signal-to-noise ratio, we identified a faint contact binary star in the target field which was chosen for a detailed analysis. In Figure\ref{chart}, we provide a finding chart of $\sim 13\times13$ arcmin target field marking the binary star at the center of the frame. We only considered those photometric data points which have an error $<$0.04 mag. In this way, we have accumulated  photometric data points for the target star on 304 and 248 frames in $R_c$ and $I_c$ bands, respectively. The data files may be obtained from the lead author or through the online data base.

The star was also identified in Magnier et al. (1992) and Massey et al. (2006) catalogues for the M31 galaxy. In recent times, photometric observations of the field containing the target star have been carried out under the Catalina Surveys (Drake et al. 2014) which reported it as a binary star. However, a detailed investigation of the photometric solutions, spectral type and period analysis of the system has not been carried out until now.
\subsection{Orbital period investigation}
\begin{figure}[h]
\centering
\vspace{0.5cm}
\includegraphics[width=14.0cm, height=7.0cm]{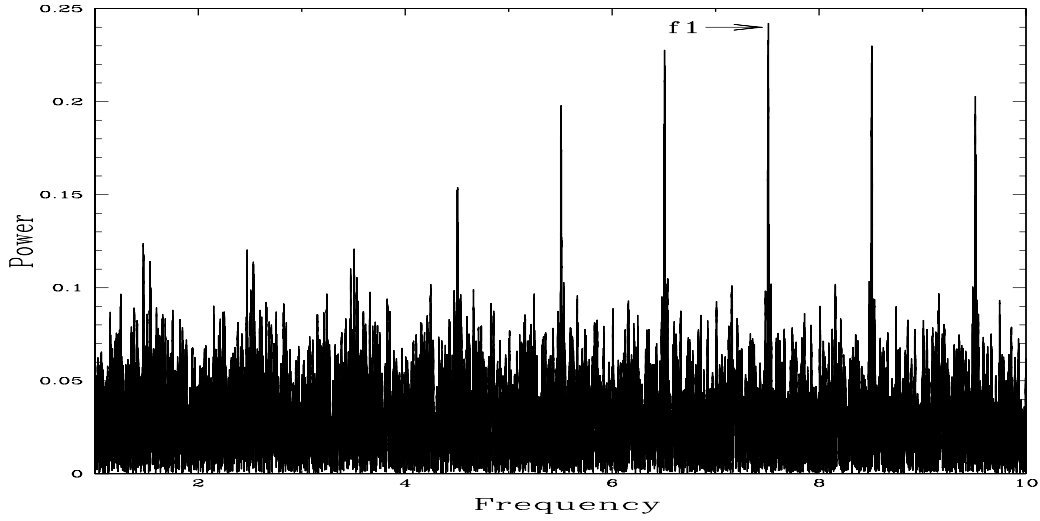}
\begin{flushleft}
\caption{Power spectra obtained from the $R_c$ band data of the star.}
\label{fourier}
\end{flushleft}
\end{figure}

The determination of accurate orbital period is very important to measure various physical parameters of the binary system as well as to understand the physical processes such as mass transfer and magnetic activity. Drake et al. (2014) first reported this star as a W UMa contact binary having a period of 0.266398 days and reported a $V$ band amplitude of 0.54 mag. In our analysis, we carried out period analysis using Period04 (Lenz \& Breger 2005). The resulting power spectrum is shown in Figure~\ref{fourier} where the peak frequency is found to be at 7.507470 day$^{-1}$. This corresponds to a period of 0.133201 day. The uncertainty in the period determination is found to be 0.000009 day. Since binary light curves are normally represented by two sine waves, therefore, multiplying the period by a factor of two gives the true period of the binary system. We therefore assigned a period of 0.266402$\pm$0.000018 days for the binary system, which is in accordance with the period obtained by Drake et al. (2014).
%
%------------------------------------------- Fig02-------------------------
%------------------------------------------- Fig02------------------------- 
%

In the following equation, we draw the ephemeris for the minimum light as a function of the binary star.
$$
Min~I (HJD) = 2451474.885657(24) + 0.266402(18) \times E
$$
The initial epoch is taken as time of minimum light in our data which was determined using the technique proposed by Kwee \& Van Woerden (1956). In the above equation, parentheses values represent the error in terms of last quoted numbers. 

\begin{figure*}
\centering
\vspace{0.5cm}
\includegraphics[width=14.0cm, height=12.0cm]{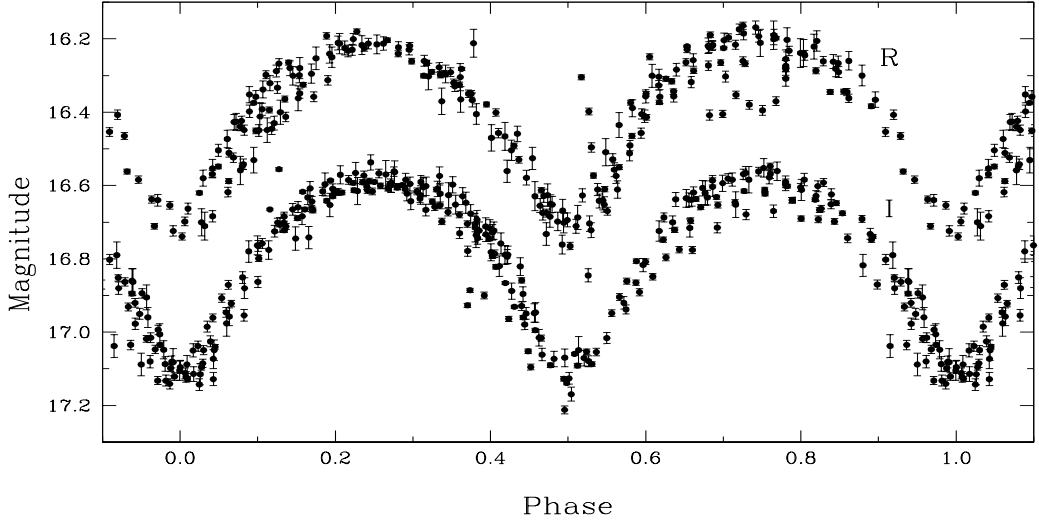}
\vspace{-5.0cm}
\begin{flushleft}
\caption{Folded $R_c$ and $I_c$ bands phase light curves for the W UMa binary.}
\label{lc}
\end{flushleft}
\end{figure*}

We determined the phase corresponding to each observation by using the estimated period. Figure~\ref{lc} illustrates the phase folded light curves of the star in both $R_c$ and $I_c$ bands. The broadening in the phase folded light curves indicate that the period is slightly changing over several years baseline of the survey. The change in period with time occur either due to redistribution of matter between the binary components, or when angular momentum is gained or lost by the system due to interaction with a third body or collision within a dense stellar environment. In our observations, we could identify only three minima which is insufficient to draw any firm conclusion on the state of binary system and further high-precision photometric observations will be needed to determine the rate of period change.
%
%------------------------------------------- Fig03-------------------------
%------------------------------------------- Fig03-------------------------
%
\subsection{Mean Magnitudes}\label{mm}
In the present analysis, we prefer to determine phase weighted mean magnitude of the binary star applying the equation given by Saha et al. (1994):
$$ \overline\sl{m} = -2.5\sl\log_{10} \mathrm{\sum_{i=1} ^{n}}\ 0.5(\phi_{i+1}-\phi_{i-1}) 10^{-0.4 m_{i}} $$
where n is number of observations and $\sl\phi_{i}$ is the phase of the $i^{th}$ data point after sorting out the period in increasing phase. The equation requires non-existent entities $\sl\phi_{0}$ and $\sl\phi_{n+1}$ which are set identical to $\sl\phi_{n}$ and $\sl\phi_{1}$, respectively. We determined a mean magnitude of 16.867 mag and 16.470 mag in $R_c$ and $I_c$ bands, respectively. The mean magnitude estimation through the phase-weighted method was preferred as this method minimizes the systematic biases due to loss of measurements towards fainter magnitudes (Saha \& Hoessel, 1990). The $UBVRI$ magnitude of this star is given as 17.620, 17.070, 17.778, 16.706 and 16.262 mag, respectively in the Massey et al. (2006) catalogue, while mean $V$ magnitude is given as 16.92 in Drake et al. (2014). The 2MASS JHK magnitudes are given as 14.345, 14.009, and 13.847 mag, respectively (Cutri et al. 2003). From the phase light curves, the amplitudes were estimated as 0.67 and 0.58 mag in $R_c$ and $I_c$ bands, respectively. The phase weighted mean magnitudes give a colour of $(R-I)_c = 0.40$ mag for the binary system. The brief summary of the physical parameters obtained for the W UMa binary $CSS\_J004259.3+410629$ is given in Table~2.
\input{table02.tex}

\subsection{Spectral Classification}
From $UBVRI$ photometric observations in the direction of M31, Massey et al. (2006) determined a colour $(B-V)=0.708$ mag for this star. From the Galactic extinction map of Schlafly \& Finkbeiner (2011), the Galactic absorption $A_V$ is found to be 0.170 mag in the direction of our target field, resulting in a reddening of $E(B-V)=0.055$ mag assuming a normal reddening law. This yields an un-reddened colour of $(B-V)_0 = 0.653$ mag for the star. According to the Cox (2000) colour index - spectral type calibration given in Table~15.7, this value of colour index corresponds the spectral type of G3V for the system.
\subsection{Parameter relationships and characterization}
Assuming that the formation of contact binary system were happened through almost normal-hydrogen core burning stars where mass-radius relation for the main-sequence stars are obeyed, Wang (1994) gave the period-colour relation as following
$$
(B-V)_0 = 0.077 - 1.003~\sl{logP}
$$
Taking the orbital period $P = 0.266402$ days for the binary system, it gives an intrinsic colour of $(B-V)_0 = 0.653$ mag which exactly matches with our photometric estimation.

The effective temperature of the primary component of the binary system can be estimated using the relation given by Wang (1994):
$$
\sl{logT_{eff}} = 3.970 - 0.310~(B-V)_0
$$
This gives an approximate temperature of 5856 K which corresponds to the spectral class of G1/2 for the primary component of the binary system. This closely matches with our previous estimation from the photometric observations of this star.

From the 2MASS catalogue, the infra-red colour index $(J-K)$ for the star turns out to be 0.50 mag. Deb \& Singh (2011) recently analysed light curves of 62 binary stars, of which most of them were contact binaries, and found a good correlation between the period and infra-red $(J-K)$ colour as following
$$
(J-K) = (0.11\pm0.01)P^{-1.19\pm0.08}
$$
where P is the period in days. When we determined $(J-K)$ using this relation for the estimated period of the binary star, we found a colour of $(J-K) = 0.53\pm0.06$ which is in very close agreement with the infra-red colour of 0.50 mag obtained from the 2MASS survey data. This does not only show that their relationship is well constrained but also that the characterization of our star as a contact binary is genuine. In order to estimate the absolute magnitude ($M_V$), we used the PLC relation given by Ruci\'{n}ski \& Duerbeck (1997) for W UMa binaries:
$$M_V = -4.44~\sl{logP} + 3.02~(B-V)_0 + 0.12$$
For the star having $V=16.92$ mag and $(B-V)_0=0.653$ mag, this gives an absolute magnitude of 4.64$\pm$0.22 mag where the error represents mean error in the relation given by Ruci\'{n}ski \& Duerbeck (1997). This resulted in an apparent distance modulus of 12.28$\pm$0.22 mag for the star. Considering the Galactic absorption $A_V = 0.17$ mag in the direction of target field from the extinction map of Schlafly \& Finkbeiner (2011), we determined an absolute distance modulus of 12.11$\pm$0.22 mag. This led to a stellar distance of about 2.64$\pm$0.03 kpc which suggests that this W UMa binary is actually a distant foreground star in the direction of M31.
\section{Photometric light curve}\label{analysis}
\begin{figure*}
\centering
\includegraphics[width=14.0cm, height=12.0cm]{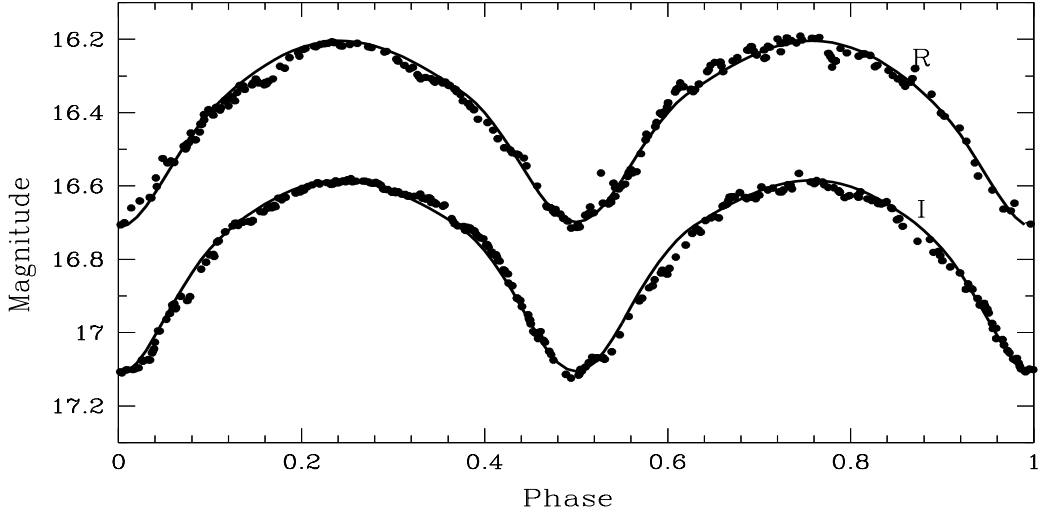}
\vspace{-5.0cm}
\begin{flushleft}
\caption{Light curves of star in $R_c$ and $I_c$ band. The continuous lines show the best fits derived through the WD code.}
\label{model}
\end{flushleft}
\end{figure*}

As observations were taken for a long period of time spanning over a few years in different observing conditions, the estimated phase values may have some errors. Therefore, we binned the data in the width of 0.02 in phase and mean magnitude and errors were determined in each phase bin. The binned phase light curve has a smaller scatter allowing a better visual identification of the binary system as shown in Figure~\ref{model}. The light curves are very symmetric and show typical EW-type variations, where depths of both minima are almost similar, which enable us to determine reliable photometric parameters of the binary system.
\subsection{q-parameter estimation}
When radial velocity measurements are available, the mass ratio between components $(q = \frac{M_2}{M_1})$ in a binary system is simply inversely proportional to the measured peak radial velocity ratio of the two components. However, if spectroscopic data is not available, a unique solution for $q$ cannot be obtained. The reason behind this is that multiple common minimum residuals in the photometric data can be obtained for a variety of parameters including $q$. However, employing an iterative approach, it is possible to restrict the $q$ within valid range where few parameters are adjusted for each value of $q$. Then it could be possible to establish $q$ which yields the lowest value of residual. As our target star is too faint ($\sim$ 17 mag) to have any spectroscopic observations, we determined $q$ from the photometric data through the $q$-search method using the Wilson and Devinney (WD)-v2003 code (Wilson \& Devinney, 1971). Here, we opted mean effective temperature of the secondary $T_2$, monochromatic luminosity of the primary $L_1$, surface potentials of the two components $\Omega_1$ = $\Omega_2$, and orbital inclination $i_0$ as flexible parameters until a convergent solution was found.

%------------------------------------------- Fig04-------------------------
\begin{figure}[h]
\centering
\includegraphics[width=8.3cm, height=10.0cm]{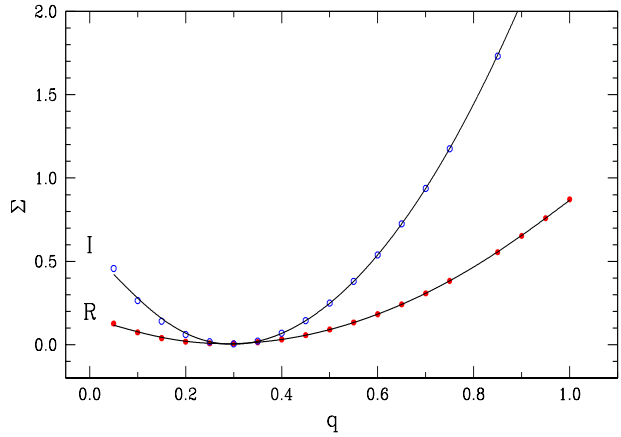}
\vspace{-4.5cm}
\begin{flushleft}
\caption{The variation in the residuals $\Sigma W(O-C)^2$ as a function of $q$. The solid points and open circles represent the data in $R_c$ and $I_c$ bands, respectively.}
\label{res}
\end{flushleft}
\end{figure}
%------------------------------------------- Fig04-------------------------

We estimated the sum of squares of the residuals $\Sigma W_i(O-C)^2_i$ with different $q$ value starting from 0.05 and in increments of 0.05. The variation of $\Sigma W_i(O-C)^2_i$ for all the assumed values of $q$ is shown in Figure~\ref{res}. An arbitrary smooth fit is also drawn in the figure to connect the data points. The minimum value of $\Sigma$ yields the best value of $q$ which is found to be $0.28\pm0.01$ and $0.29\pm0.01$ in $R_c$ and $I_c$ bands, respectively. Hence we take $q=0.28$ as the initial value.
%
%
%------------------------------------------- Fig05-------------------------
%----------------------------------------------------------
%
\subsection{Modeling and photometric solutions}
To examine the geometrical structure and evolutionary state of the binary system, we analyzed the photometric data using the WD code (Wilson \& Devinney 1971, Wilson 1990, 1994). As explained in the earlier section, the mass ratio (q) of the binary is fixed at q = 0.28.  While modeling the photometric light curves, mode 3 for the contact system is chosen. The gravity darkening coefficients and bolometric albedo adoptable for convective envelopes are taken, which were $g_1$ = $g_2$ = 0.32 (Lucy 1967), $A_1$ = $A_2$ = 0.5 (Ruci\'{n}ski 1969), the limb-darkening coefficients $x_1$ = $x_2$ = 0.6 (Al-Naimiy, 1978), $F_1=1$ and $IFAT_1 = 1$. The eccentricity $e$ of the orbit for the contact binary was assumed as 0. Further, we assumed third light $l_3 = 0$ and the longitude of periastron $\omega = 90^0$. In order to compute $L_2$ from $T_1$, $T_2$, $L_1$ and the radiation laws, the control integer $IPB$ was assigned as 0. The model fits were carried out as explained in Joshi et al. (2016). The results of our best fit solution are given in columns 2 and 3 of Table~3 for the $R_c$ and $I_c$ bands, respectively. The $R_c$ and $I_c$ band light curves along with the model fits, are shown in the Figure~\ref{model}. The fit of the theoretical curves seems to match well, despite scatter in the photometric data. The mass ratio and fill out factor along with the small difference in temperature of $\Delta(T_1 - T_2)  \sim 120$ K suggest that the star is a A-sub-type deep contact binary system, implying that the deeper eclipse is the transit of the massive star by the less massive one. 
\input{table03.tex}
\subsection{Geometric solutions}
\begin{figure*}
\centering
\vspace{0.5cm}
\includegraphics[width=14.7cm, height=11.0cm]{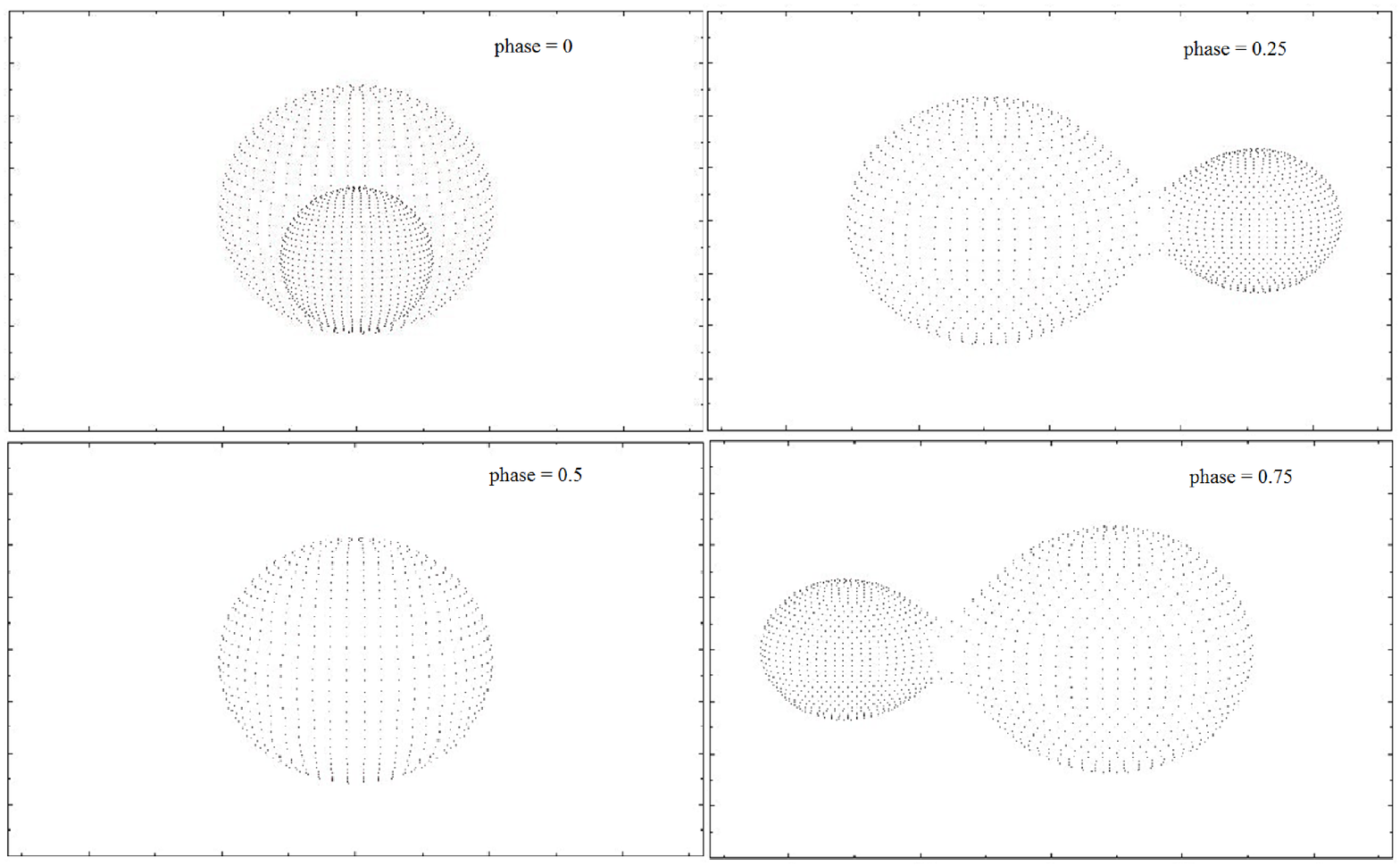}
\begin{flushleft}
\caption{Geometric configurations for the binary star at phase 0.0, 0.25, 0.50, and 0.75.}
\label{geo}
\end{flushleft}
\end{figure*}

Considering fill out factors of 47\% and 44\% in the $R_c$ and $I_c$ bands, respectively and temperature difference of about 120 K between the two components, one can assume a good thermal contact in binary system and both the components are stable and least likely to merge (e.g., Kahler, 2004). The inclination angle was determined to be $\sim 77^o$ for the binary star, indicating the eclipsing nature of the star. A usual approach of describing binary systems is through the Roche model. In Figure~\ref{geo}, we show the geometrical configurations of the binary star at phases 0.25, 0.50, 0.75 and 1.0. From the photometric light curves, we have not seen any O'Connell effect in the binary star which, in general, is linked with the existence of cool spots on any component of the binary system (Kalomeni et al., 2007).

%
%------------------------------------------- Fig06-------------------------
%------------------------------------------- Fig06-------------------------
%

\subsection{Absolute parameters}\label{abs_par}
Due to the lack of radial velocity (RV) solutions, we used empirical relations to determine the absolute parameters of the binary system. Dimitrov \& Kjurkchiva (2015) gave a period--semi-major axis (P-a) relation on the basis of 14 binary stars having $P < 0.27$ d which had both RV and photometric solutions, which is approximated by a parabola:
$$
a = -1.154 + 14.633~P - 10.319~P^2
$$
where P is in days and a is in solar radii. Following the above relation, we
determined a semi-major axis $a(R\odot) = 2.01\pm0.05$.

 Following Newton's formulation of Kepler's third law we can calculate the system's total mass as 
$$ M_1+M_2 = a^3/P^2$$
where $P$ is in yr and $a$ in AU. It gives a total mass of $1.52\pm0.09 M_\odot$ for the binary system. According to the light curve solution in the $R_c$ band, the mass ratio $q$ of the system is found to be $0.28\pm0.01$ which gives a mass of $1.19\pm0.09$ and $0.33\pm0.02$ for the primary and secondary components, respectively. The mean fractional radii of the components were obtained with the formula
$$ r_{mean} = (r_{pole} \times r_{side} \times r_{back})^{1/3}$$
It gives mean fractional radii of $0.51\pm0.02$ and $0.29\pm0.04$ for the primary and secondary components, respectively.  Using the semi-major axis, we can calculate the radii of the binary components as $R = a \times r_{mean}$ which gives radii of $R_1 = 1.02\pm0.04 R_\odot$ and $R_2 = 0.58\pm0.08 R_\odot$. Since the sum of mean fractional radii of the binary components is $0.80\pm0.04$ which is greater than 0.75, the system is expected to be in the state of good thermal contact (Kopal, 1959) as we earlier found.

The absolute parameters of bolometric magnitudes and luminosity can be calculated using the equations
$$ M_{bol} = 4.77 - 5~\sl{log(R/R_\odot)} - 10~\sl{log(T/T_\odot)} $$ 
$$ L = (R/R_\odot)^2~(T/T_\odot)^4$$
Considering a solar temperature of $T_\odot=5780$ K, we found bolometric magnitudes of $5.17\pm0.19$ and $6.31\pm0.79$ for the primary and secondary components and their respective Luminosities are found to be $0.69\pm0.05~L_\odot$ and  $0.24\pm0.08~L_\odot$. The mean densities of the binary components were derived from the following equation given by Mochnacki (1981):
$$ \rho = \frac{0.0189}{r_{mean}^3P^2~(1+q)}$$
The mean densities of the primary and secondary components are estimated as $1.61\pm0.19$ and $2.50\pm1.03$, respectively. A similar but independent analysis was also done for the $I_c$ band data and we found almost similar values in both the bands. A summary of the absolute parameters of the binary system in both $R_c$ and $I_c$ bands are given in Table~4.

\input{table04.tex}
\subsection{Mass-Radius relation}\label{mass_rad}
\begin{figure}[t]
\centering
\includegraphics[width=8.3cm, height=11.6cm]{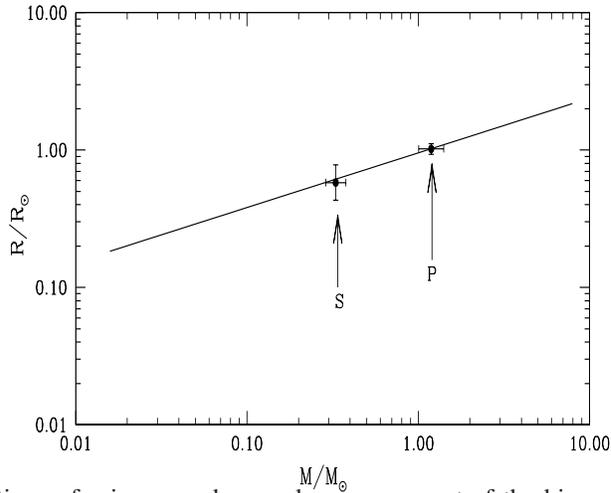}
\vspace{-5.4cm}
\begin{flushleft}
\caption{The positions of primary and secondary component of the binary on a mass-radius diagram. The continuous line shows the zero-age main sequence from Schmidt-Kaler (1982).}
\label{mass-rad}
\end{flushleft}
\end{figure}

The absolute parameters of binary components given in Table~4 are used to draw the evolutionary status of the system. For this we have drawn the mass and radius of both the components in the mass-radius diagram shown in Figure~\ref{mass-rad}. The continuous line represents the theoretical zero-age main sequence (ZAMS) taken from Schmidt-Kaler (1982) which is defined by the following relation:
$$
\sl{log(R)} = 0.917~\sl{log(M)} - 0.020 ~~~~~~~~~~ (-1.0<\sl{log(M)}<0.12)
$$
Both the components fall exactly over the ZAMS which suggests that W UMa binary lies in the main-sequence. It also means that the mass transfer between the components is not significant enough to have resulted in restructuring the secondary component at this stage.
%
%--------------------- Fig07-------------------------
%--------------- End Fig 07-------------------------

\section{Discussion}\label{discuss}
\begin{figure}
\centering
\includegraphics[width=8.3cm, height=7.0cm]{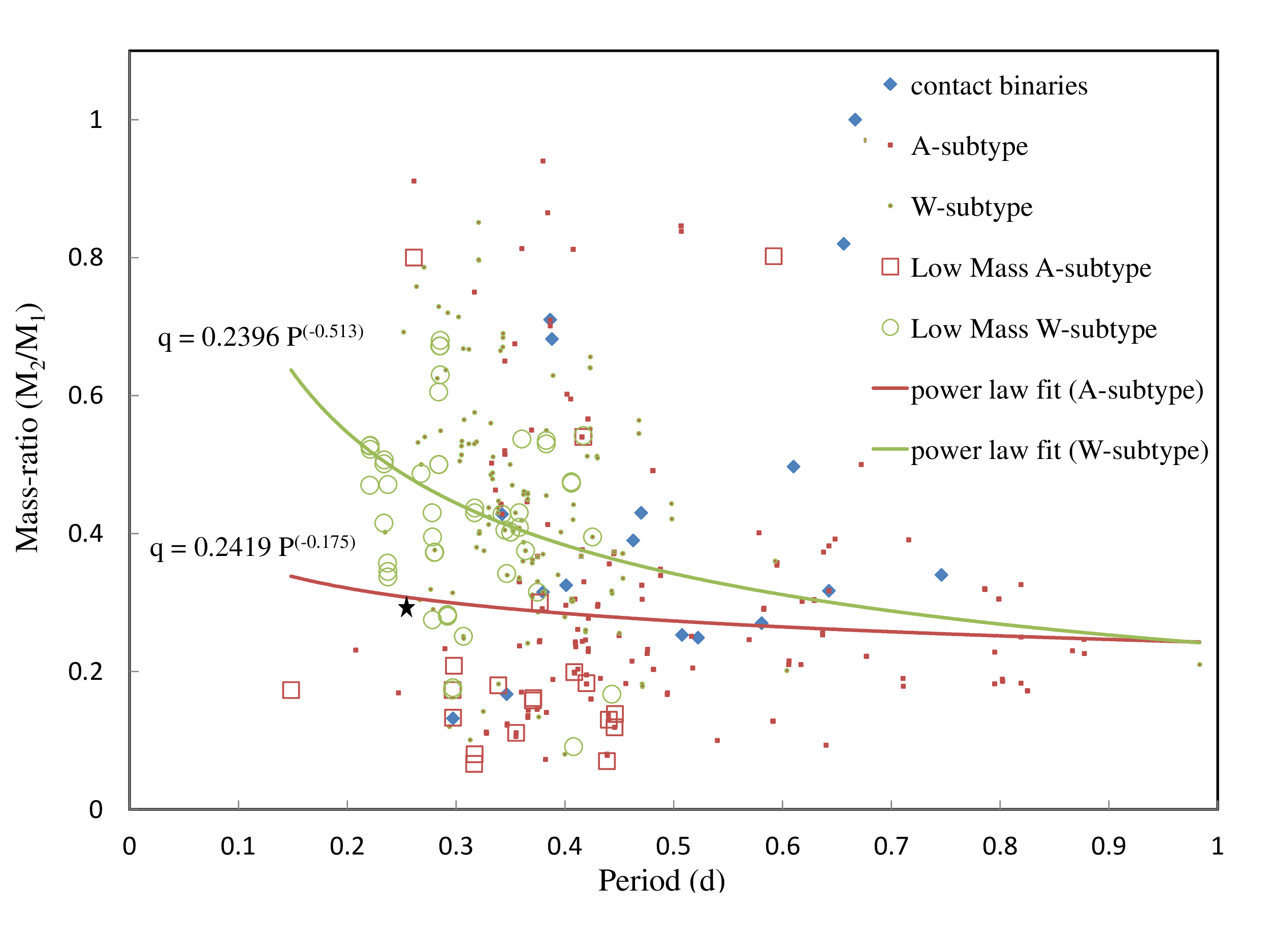}
\vspace{-0.3cm}
\begin{flushleft}
\caption{Mass-ratio vs Period plotted for $\sim$ 300 contact binaries including low mass binaries in A-subtype and W-subtype. The star in black colour represents the binary system $CSS\_J004259.3+410629$.}
\label{mass_ratio}
\end{flushleft}
\end{figure}

It is observed that A-subtype contact binaries are dominant for $q<0.25$, while W-subtype contact binaries exist in the early stage of evolution with $q>0.4$ and both the subtypes are equally common in the region $0.25<q<0.4$ (Yildiz \& Dogan, 2013). The energy transfer in the common convective envelope of the contact binaries may occur in the deeper layers for A-subtype but in the outermost layers for W-subtype. Lifang Li et al. (2004) have discussed a model for contact binaries with a low total mass, where binary star evolves into a system with smaller mass-ratio and deeper envelope and concluded that some of the A-subtype contact binaries observed with low total mass could be W-subtype binaries in their later evolutionary stages. In order to understand any such transition in the presently studied W UMa binary, we draw mass-ratio versus period for the binary systems in Figure~\ref{mass_ratio}, and also over-plotted the parameters of $\sim$ 300 contact binaries  detected in several previous studies including many low mass A-subtypes and W-subtypes W UMa binaries (Pribulla et al., 2003; Deb \& Singh, 2011; Eker et al., 2013; Yildiz \& Dogan, 2013). The position of the binary system $CSS\_J004259.3+410629$ in the plot indicates its possible transition from W-subtype to A-subtype while undergoing a decrease in mass-ratio via magnetic braking through angular momentum loss and it may continue to evolve into a single star through merger. A secular increase in the period of the system is normally detected in the systems evolving to higher fill-out factors and lower mass-ratios before reaching to the stage of merger as originally proposed by Soker \& Tylenda (2003). In a most recent observational evidence of such merger scenario, Tylenda et al. (2011) and Zhu et al. (2016) has reported a merger of contact binaries of V1309 Sco into a single object resulting an eruption in the star due to the ejection of the primary envelope.
%
%--------------------Fig08---------------------------
%-------------------------End Fig08---------------------------

%
\section{Results and Conclusions}\label{conclu}
W UMa variables are over-contact eclipsing binary systems. In order to understand how these systems form and evolve requires photometric and spectroscopic observations spanning many years, followed by detailed analysis through accepted models. In the present study we have carried out a detailed photometric analysis of a recently discovered W UMa binary system $CSS\_J004259.3+410629$ falling in the field of M31 which was observed during the Nainital Microlensing Survey. Our analysis yields a period of $0.266402\pm0.000018$ days for the binary system. The photometric and infra-red data suggest a spectral type of G3V for the binary system. The photometric data was examined through the WD-code which supplied the mass ratio between the two components as $0.28\pm0.01$. The more massive component is found to have slightly higher surface temperature than the less massive one. The mass ratio and fill out factor along with the small temperature difference between the two components suggest that the star is a A-subtype deep contact binary system. The absolute parameters of the binary components are deduced using various empirical relations which yield mass, radius and luminosity of 1.19$\pm$0.09, 1.02$\pm$0.04, 0.69$\pm$0.05  and 0.33$\pm$0.02, 0.58$\pm$0.08, 0.24$\pm$0.08, respectively for the two components. Along with its position in the mass-radius diagram, the values of mass, radius and luminosity deduced for the massive primary component closely agree with the G3 spectral class of the main sequence star which was also ascertained from the photometric data. The distance between the two components is estimated as $a$ = 2.01 $R_\odot$, according to the third Kepler's law. To understand the properties of the binary system and variation in its orbital period, further long-term photometric and spectroscopic monitoring of this system would be required.

\normalem
\begin{acknowledgements}
YCJ acknowledges the financial support from the project DST/INT/SA/P-02 and RJ acknowledges the financial support from the project UGC-BSR research Start-Up Grant Sanctioned vide UGC Order No. F. 30-108/2015(BSR) of UGC, under which part of the work has been carried out. We acknowledge Chris Engelbrecht for reading the draft carefully which further improves the paper.
\end{acknowledgements}

\end{document}

%% file: table01.tex
 \begin{table*}
    \caption{Detail of the observations taken under Nainital microlensing survey}
    \begin{center}
\begin{tabular}{ccccccc}
\hline \\
   Date and image ID  &       JD           &  FWHM    & Filter &  Exp    &    Tel     &  CCD          \\
  yyyymmdd            &                    & (arcsec) &        & (sec)   &            &               \\ 
\hline    \\                                           
19981121i01           &   2451139.204757   &   1.83   &   $I_c$    &  1200   &    104-cm  &  $1k\times1k$   \\
19981121i02           &   2451139.219572   &   2.18   &   $I_c$    &  1200   &    104-cm  &  $1k\times1k$   \\
19981121r01           &   2451139.157292   &   1.76   &   $R_c$    &  1200   &    104-cm  &  $1k\times1k$   \\
     .                &         .          &    .     &     .      &    .    &     .      &      .          \\
     .                &         .          &    .     &     .      &    .    &     .      &      .          \\
20041027i01           &   2453306.244826   &   2.10   &   $I_c$    &    60   &    200-cm  &  $2k\times2k$   \\
20041027i02           &   2453306.246817   &   2.60   &   $I_c$    &    60   &    200-cm  &  $2k\times2k$   \\
     .                &         .          &    .     &     .      &    .    &     .      &      .          \\
     .                &         .          &    .     &     .      &    .    &     .      &      .          \\
20111102r05           &   2455868.138426   &   2.11   &   $R_c$    &   300   &    104-cm  &  $2k\times2k$   \\
20111102r06           &   2455868.142951   &   1.99   &   $R_c$    &   300   &    104-cm  &  $2k\times2k$   \\ \\
\hline                                               
\end{tabular}                                        
\end{center}                                         
    \label{Tab1}
\end{table*}

%% file: table02.tex
 \begin{table*}
    \caption{Details of the W UMa binary star under study}
    \begin{center}
\begin{tabular}{cccccccc}
\hline \\
ID &    RA   & DEC  & $V$  & $R$  &  $\Delta$R & $(R-I)$ & Period   \\
   & (J2000) & (J2000) &  (mag) & (mag) &  (mag)     & (mag)     & (day)    \\
\hline  \\
ID10853 & 0:42:59.33 & +41:06:29.3  & 16.920 & 16.867  & 0.67     & 0.40  & 0.266402  \\ \\
\hline
\end{tabular}
\end{center}
    \label{Tab1}
\end{table*}

%% file: table03.tex
 \begin{table}
    \caption{The Photometric parameters obtained for the W UMa binary star using the WD method in
    both $R$ and $I$ band.}
\small
    \begin{center}
\begin{tabular}{lccccc}
\hline \\
Elements                        &&    $R$ band               &&     $I$ band          \\
\hline \\
$T_1$ (K)                       && 5724		             &&  5724	              \\
$T_2$ (K)		        && 5706 $\pm$ 55	     &&  5786 $\pm$ 58	      \\
Spectral Type			&& G3    		     &&  G3	              \\ 
$q$		                && 0.28 $\pm$ 0.01	     &&  0.29 $\pm$ 0.01       \\
$i_0$		                && 77.3 $\pm$ 0.4	     &&  79.3 $\pm$ 0.6       \\
$\Omega_{1,2}$			&& 2.38 $\pm$ 0.06           &&  2.38 $\pm$ 0.01     \\
$f$			 	&& 46.70\%		     &&  44.10\%              \\
$r_1$        Pole 		&& 0.47 $\pm$ 0.01	     &&  0.47 $\pm$ 0.01      \\
             Side 		&& 0.51 $\pm$ 0.02	     &&  0.51 $\pm$ 0.01      \\
             Back 		&& 0.54 $\pm$ 0.03	     &&  0.54 $\pm$ 0.01      \\
$r_2$        Pole		&& 0.26 $\pm$ 0.03	     &&  0.27 $\pm$ 0.01      \\
             Side		&& 0.28 $\pm$ 0.03	     &&  0.28 $\pm$ 0.01      \\
             Back		&& 0.32 $\pm$ 0.07	     &&  0.33 $\pm$ 0.01      \\
$L_1$				&& 0.761		     &&  0.745	              \\
$L_2$				&& 0.239		     &&  0.255	              \\
$L_3$				&& 0.0 		             &&  0.0 	              \\
$x_1$				&& 0.6 		             &&  0.6 	              \\
$x_2$				&& 0.6 		             &&  0.6 	              \\
$A_1$				&& 0.5 		             &&  0.5 	              \\
$A_2$				&& 0.5 		             &&  0.5 	              \\
$G_1$				&& 0.32		             &&  0.32	              \\
$G_2$				&& 0.32		             &&  0.32	              \\
$\sigma$			&&0.0058		     &&  0.0027		      \\  \\
\hline                                                                               
\end{tabular}                                                                                                 
\end{center}						                             			 
    \label{Tab2}					                             			 
\end{table}						                             			 

%% file: table04.tex
\begin{table}
\caption{The fundamental parameters obtained for the W UMa binary star. Here, subscript "1" is used for higher-mass component and "2" for the lower-mass component of the binary system. All values are given in the solar units.}
    \begin{center}
\begin{tabular}{lcccc}
\hline \\
Parameters&&     $R$ band     &&  $I$ band  \\
\hline     \\                                                
$M_1~(M_\odot)$    &&  1.19 $\pm$ 0.09 &&   1.18 $\pm$ 0.09   \\
$M_2~(M_\odot)$    &&  0.33 $\pm$ 0.02 &&   0.34 $\pm$ 0.03   \\
$R_1~(R_\odot)$    &&  1.02 $\pm$ 0.04 &&   1.02 $\pm$ 0.04   \\
$R_2~(R_\odot)$    &&  0.58 $\pm$ 0.08 &&   0.59 $\pm$ 0.02   \\
$L_1~(L_\odot)$    &&  0.69 $\pm$ 0.05 &&   0.69 $\pm$ 0.05   \\
$L_2~(L_\odot)$    &&  0.24 $\pm$ 0.08 &&   0.26 $\pm$ 0.03   \\
$M_{bol1}$         &&  5.17 $\pm$ 0.19 &&   5.17 $\pm$ 0.19   \\
$M_{bol2}$         &&  6.31 $\pm$ 0.79 &&   6.25 $\pm$ 0.28   \\
$\rho_1$ (cgs)     &&  1.61 $\pm$ 0.19 &&   1.59 $\pm$ 0.19   \\
$\rho_2$ (cgs)     &&  2.50 $\pm$ 1.03 &&   2.40 $\pm$ 0.25   \\
$a (R_\odot)$      &&  2.01 $\pm$ 0.05 && \\
$d$ (kpc)          &&  2.64 $\pm$ 0.03 && \\ \\
\hline
\end{tabular}
\end{center}
\label{Tab4}
\end{table}